\documentclass{article}
\usepackage{spconf,amsmath,graphicx}
\usepackage{booktabs}
\usepackage{float}
\usepackage{cite}
\usepackage[T1]{fontenc}
\usepackage{amsfonts}
\usepackage{bm}
\usepackage{algorithm}
\usepackage[noend]{algpseudocode}
\usepackage{amsmath}
\usepackage{amsfonts}
\usepackage{multirow}
\usepackage{multirow}
\usepackage{soul}
\usepackage{multicol}
\title{An adapter based multi-label pre-training for speech separation and enhancement}
\name{Tianrui Wang$^{1}$, Xie Chen$^{2,3}$, Zhuo Chen$^{4}$, Shu  Yu$^{2}$, Weibin Zhu$^{1}$}
\address{$^{1}$Institute of Information Science, Beijing Jiaotong University, Beijing, China\\
	$^{2}$MoE Key Lab of Artificial Intelligence, AI Institute, \\Shanghai Jiao Tong University, Shanghai, China\\
	$^3$ Peng Cheng Laboratory, Shenzhen, China \\
	$^{4}$Microsoft, One Microsoft Way, Redmond, WA, USA
}

\begin{document}
	%\ninept
	%
	\maketitle
	\begin{abstract}
		In recent years, self-supervised learning (SSL) has achieved tremendous success in various speech tasks due to its power to extract representations from massive unlabeled data. However, compared with tasks such as speech recognition (ASR), the improvements from SSL representation in speech separation (SS) and enhancement (SE) are considerably smaller. Based on HuBERT, this work investigates improving the SSL model for SS and SE. We first update HuBERT's masked speech prediction (MSP) objective by integrating the separation and denoising terms, resulting in a multiple pseudo label pre-training scheme, which significantly improves HuBERT's performance on SS and SE but degrades the performance on ASR. To maintain its performance gain on ASR, we further propose an adapter-based architecture for HuBERT's Transformer encoder, where only a few parameters of each layer are adjusted to the multiple pseudo label MSP while other parameters remain frozen as default HuBERT. Experimental results show that our proposed adapter-based multiple pseudo label HuBERT yield consistent and significant performance improvements on SE, SS, and ASR tasks, with a faster pre-training speed, at only marginal parameters increase.
		
	\end{abstract}
	\section{Introduction}
	% 这一段主要是引子
	Self-supervised learning (SSL) has recently become a fast-growing topic in speech processing \cite{apc, cpc}. Different from supervised learning, the supervision of SSL is derived from the data itself, which allows the models to leverage massive unlabeled data and thus achieve significant improvements in various tasks \cite{tera, wav2vec, vqwav2vec, wav2vec2, hubert, data2vec}. A typical SSL approach for speech processing includes two stages. Firstly, an upstream model is pre-trained based on the unlabeled data. Secondly, a small set of supervised data is employed to fine-tune the network for each downstream task by updating the entire upstream model or adding task-specific layers \cite{SSLapplication}. In this way, each downstream task can be benefited from the extended data scale, resulting in improved performance. The SSL systems demonstrated their clear performance advantages over the supervised baselines in the SUPERB \cite{superb} and SUPERB-SG \cite{superbsg} benchmarks that cover evaluations for 15 speech tasks.
	
	However, compared with other speech tasks, the SSL models perform mediocrely on speech separation (SS) and enhancement (SE) \cite{superbsg}. One potential reason for this performance discrepancy lies in the mismatches between the pre-training scheme and their fine-tuning targets. Although features of shallow encoder layers and the introduction of additional acoustic features have been verified to be helpful for SE \cite{acousticinforloss}, the mismatches caused by SSL objective and data configuration have not been essentially addressed. SS aims to separate overlapped speech into each speech component \cite{deepcluster}, and SE aims to extract speech from a noisy environment \cite{HGCN, HGCN+}.
	In contrast, most SSL methods are only pre-trained on clean speech with a single active speaker.
	Therefore, the \textbf{data domain mismatch} is the first barrier to SS and SE. 
	Although \cite{robustwav2vec} verified that data augmentation could reduce the data domain mismatch to improve the performance on out-of-domain ASR, no such exploration was investigated for SS and SE tasks.
	On the other hand, most SSL training targets recovering only one speaker from input audio \cite{wavlm, unispeechsat}, which could be sub-optimal for tasks such as SS, where all involved speakers are equally important. We refer to this as a \textbf{task mismatch}. 
	Moreover, considering the different modeling requirements for separation tasks (SS, SE, etc.) and semantic tasks (ASR, etc.), it is challenging for one universal SSL upstream model to achieve the optimum performance in both \cite{adapterforspeech1}.
	
	% 介绍我们的方法
	To alleviate the data domain and task mismatches, we introduce two extensions to HuBERT \cite{hubert} in this study. Firstly, a data augmentation that includes raw speech, noisy speech, overlapped speech, and noisy overlapped speech is employed for pre-training. And a multiple pseudo label MSP loss is proposed to integrate separation and denoising into pre-training. In this way, HuBERT's performances are significantly improved on SS and SE but degraded on ASR. Secondly, the adapter layers are further integrated into the proposed system to address this degradation. With the adaptor-based architecture and the multiple-label MSP, we show that satisfactory performances in all three tasks can be obtained by only updating a few new parameters from the HuBERT baseline.
	
	\section{PRELIMINARIES}
	
	\subsection{HuBERT}
\begin{figure}[htb]
		\centering
		\vspace{-0.6cm}
		\includegraphics[width=9.1cm]{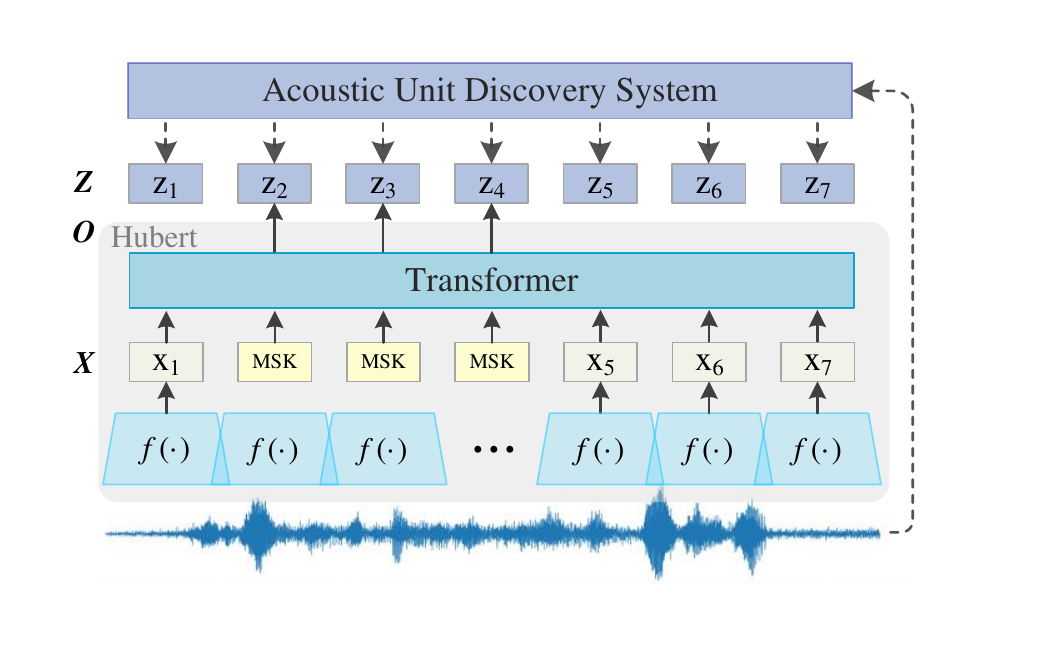}
		\vspace{-1.5cm}
		\caption{Diagram of HuBERT, which takes raw waveform as input to perform a BERT-like self-supervised pre-training.}
		\vspace{-0.3cm}
		\label{hubertpic}
	\end{figure}

	HuBERT \cite{hubert} is an SSL method which benefits from an offline clustering to generate pseudo labels $\bm{Z}=\left[z_1,z_2,\cdots,z_T\right]$, where each $z$ is a $C$-class categorical variable for a BERT-like pre-training \cite{bert}, as shown in Fig.~\ref{hubertpic}. A convolutional module (CNN) $f(\cdot)$ transforms the waveform into the frame-level feature  $\bm{X} =\left[\bm{x}_1,\bm{x}_2,\cdots,\bm{x}_T\right]$ with a stride size of 20ms. Then, the frame-level feature is encoded by the Transformer into the representation $\bm{O}=\left[\bm{o}_1,\bm{o}_2,\cdots,\bm{o}_T\right]$. During pre-training, the frame-level features are masked randomly and successively, then fed to Transformer, the model is finally trained to predict the pseudo labels of the masked frames. The distribution over codewords is parameterized as,
	\begin{equation}
		\setlength{\abovedisplayskip}{4.5pt}
		\setlength{\belowdisplayskip}{4.5pt}
		p(c|\bm{o}_t) = \frac{\exp\left(\text{sim}\left(\bm{A}\bm{o}_t,\bm{e}_c\right)/\tau\right)}{\sum_{c'}^{C}\exp\left(\text{sim}\left(\bm{A}\bm{o}_t,\bm{e}_{c'}\right)/\tau\right)}
		\label{hubert_loss}
	\end{equation}
	where $\bm{A}$ is a projection matrix, $\bm{e}_c$ is the embedding for codeword c, $\text{sim}\left(\cdot,\cdot\right)$ computes the cosine similarity between two vectors, and $\tau$ scales the logit which is set to 0.1. The prediction loss is only applied over the masked frames.
	\subsection{Adapter Based Transfer Learning}
	Adapters are lightweight modules inserted in layers of a deep neural network \cite{adapterfornlp1,adapterfornlp2}, and only the inserted adapters need to be trained for a new task. In this way, adapter-based training can retain the original capabilities of the model while being competent in a new job \cite{adapterforspeech1, adapterforspeech2}. 
	The fully connected bottleneck adapter is widely employed in BERT for various tasks \cite{adapterfornlp2}. We use these adapters for HuBERT, which will be detailed in section~\ref{adapter_hubert}.
	% Inspired by the success of the adapter employed in BERT \cite{adapterfornlp2}, the fully connected bottleneck adapters are inserted into HuBERT to be pre-trained for SS and SE in this study.
	
	\section{Proposed Method}
	Our proposed method consists of two parts. First, the multiple pseudo label based on the data augmentation is proposed to alleviate the data domain and task mismatches. Second, the adapters are employed to learn the representations for SS and SE while retaining the original task performance.

	\subsection{Multiple Pseudo Label based HuBERT}
	\label{HuBERT}
\begin{figure}[htb]
		\centering
		\vspace{-0.15cm}
		\includegraphics[width=7.5cm]{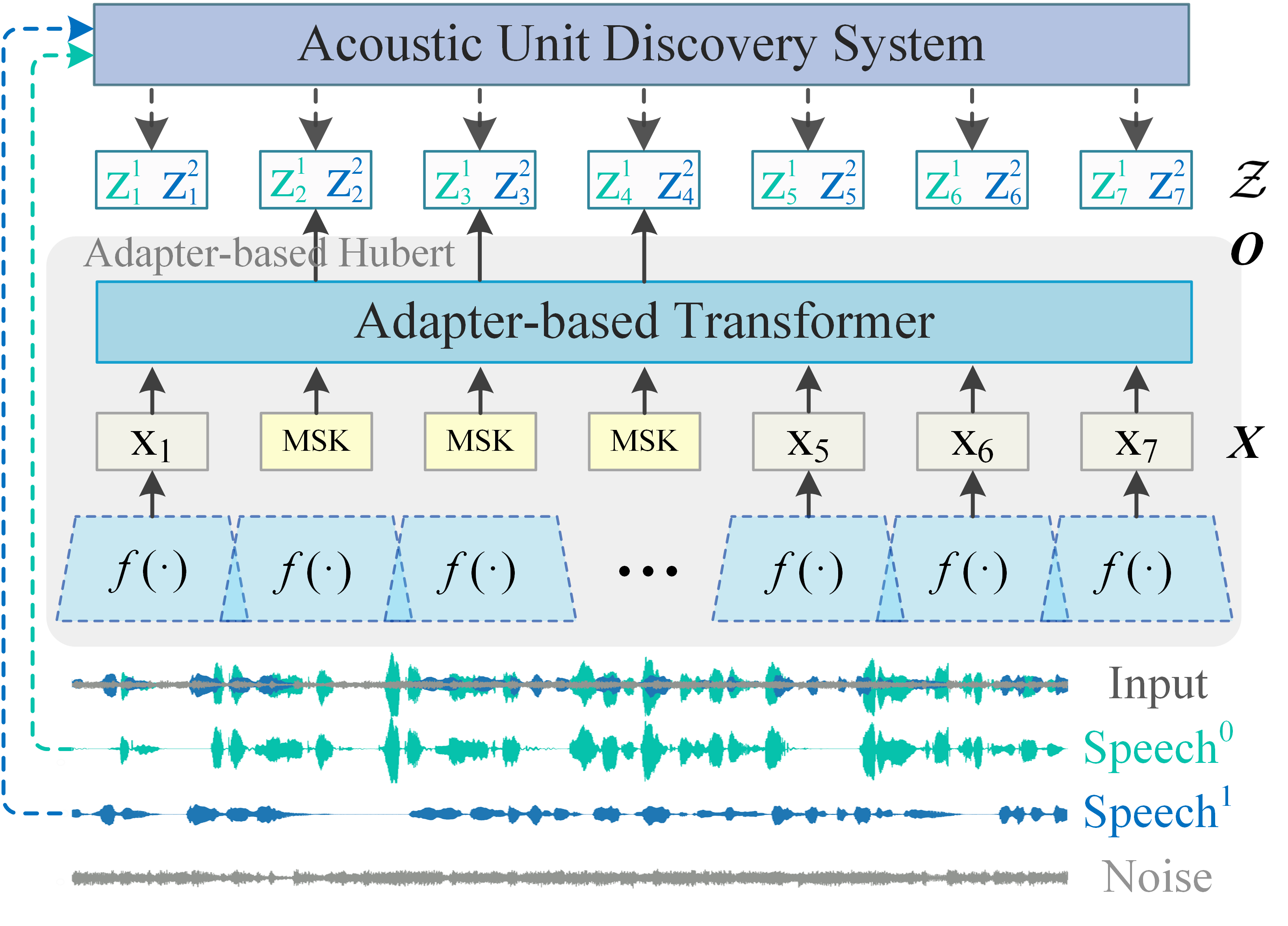}
		\vspace{-0.5cm}
		\caption{Diagram of the proposed multi-label HuBERT, which is pre-trained to predict the pseudo labels corresponding to the speech components from a given mixed input. The CNN $f(\cdot)$ with dashed and the Transformer of the original HuBERT are frozen during the adapter-based pre-training.}
		\label{multilabel}
	\end{figure}

	The proposed multiple pseudo label (multi-label) based HuBERT is pre-trained to predict the pseudo labels of speech components from the given input that includes overlapped and noisy utterances. This scheme applies to the situation of any overlapped speakers, and the 2-speaker condition is our major focus, as shown in Fig.~\ref{multilabel}. 
	
	Four kinds of audio are applied for pre-training, raw speech, noisy speech, overlapped speech, and noisy overlapped speech.
	Following HuBERT's clustering, each speech component of the mixtures is annotated with the pseudo labels $\bm{Z}$, resulting in $\mathcal{Z}\in \mathbb{R}^{N\times T}=\left[\bm{Z}_1,\bm{Z}_2,\cdots,\bm{Z}_N\right]$, where $N$ is the number of speech involved in the pre-training data.
	
	Following Eq.~(\ref{hubert_loss}), during pre-training, two series of labels $\bm{Z}^1,\bm{Z}^2$ are predicted by two pairs of codewords and projection matrix, respectively. 
	The targets of raw and noisy speech are set respectively as the label of speech and $\bm{0}$ to mimic denoising ($\bm{Z}^1=\bm{Z}_n,\bm{Z}^2=\bm{0}$). The targets of overlapped and noisy overlapped speech are set as the labels of speech components to mimic separation ($\bm{Z}^1=\bm{Z}_{n_1},\bm{Z}^2=\bm{Z}_{n_2}$), where the average energy of speech $n_1$ is greater than that of $n_2$, avoiding the permutation problem during training \cite{pit}.

	\subsection{Adapter Based HuBERT}
	\label{adapter_hubert}
\begin{figure}[htb]
		\centering
		\vspace{-1cm}
		\includegraphics[width=7cm]{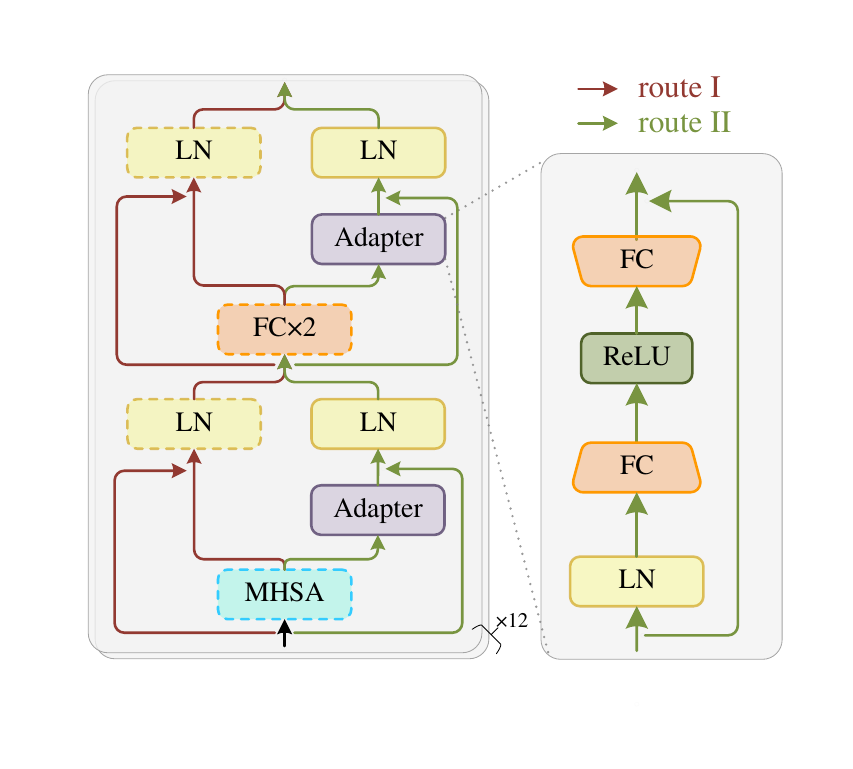}
		\vspace{-1.2cm}
		\caption{The architecture of the adapter-based Transformer blocks. Two adapters with LNs are inserted into the blocks to learn the new representations for SS and SE. The modules with dashed are frozen during pre-training. Two inferencing routes can lead to different performances during evaluation.}
		\vspace{-0.3cm}
		\label{adaptertransformer}
	\end{figure}

	Adapter-based HuBERT learns the representations for SS and SE by the inserted adapter modules, and other parameters remain frozen as default HuBERT, as shown in Fig.~\ref{adaptertransformer}. In the inference phase, route \uppercase\expandafter{\romannumeral2} with specific-task adapters can be adopted for SS and SE. Route \uppercase\expandafter{\romannumeral1} will lead to the standard HuBERT and can be applied to the original tasks, such as ASR.
	
	Our HuBERT is initialized with the open pre-trained model at first. Then the bottleneck adapter layers with near-identity initialization \cite{adapterfornlp2} are inserted after the multi-head self-attention (MHSA) and feedforward layer. The adapter consists of layer normalization (LN), fully-connect layer (FC), and ReLU, which first projects the $d$-dimensional feature into a smaller dimension with nonlinearity, then projects back to $d$ dimensions. The different features  between mixed and clean speech will substantially vary the parameters of LNs. So the adapter uses a new LN layer, which is not shared with the original task. In other words, the parameters of the original HuBERT are entirely frozen, and only the inserted adapters and LNs are pre-trained for multi-label MSP. 
	
	\section{Experiments}
	
	\subsection{Dataset}
	We conducted the experiments on the data from the LibriSpeech \cite{librispeech}, WHAM! \cite{wham}, and DNS-Challenge \cite{dns} datasets. All the audio is sampled at 16KHz.
	
	\textbf{Pre-training.} A total of 3840 hours of training and validation sets (Librispeech$960$ and three-fold simulated data) were prepared from LibriSpeech’s speech and DNS-Challenge’s noise. The ratio of raw speech, noisy speech, overlapped speech, and noisy overlapped speech was 1:1:1:1. The speech was randomly mixed with another one at a signal noise ratio (SNR) within 1dB to 6dB for an overlapped speech (the SNR represents the ratio of speech $n_1$ to speech $n_2$ mentioned in Section \ref{HuBERT}). The noise was randomly mixed with speech or overlapped speech at SNRs within -5dB to 15dB. 
	
	\textbf{SS fine-tuning} Following SUPERB-SG \cite{superbsg}, a total of 50 hours of training, validation, and test sets were simulated from LibriSpeech’s \textit{train-clean-100} and \textit{test-clean} with WHAM!’s noise. The training, validation, and test sets contained 43.3, 4.5, and 4.2 hours of speech. The condition of \textit{two speakers} and \textit{mix\_clean} was the focus of our experiment.
	
	\textbf{SE fine-tuning} Following DNS-Challenge \cite{dns}, a total of 50 hours of training and validation sets were simulated from the DNS-Challenge set. The ratio of training to validation set was 9:1, and the SNR range was -5dB to 15dB. 10 SNRs (-6dB to 21dB with an interval of 3dB) were used to simulate 5 hours of testing sets. The duration of audio clips for training and testing were 5s and 10s, respectively. 
	
	\textbf{ASR fine-tuning} The LibriSpeech's \textit{train-clean-100}, \textit{dev-clean}, and \textit{test-clean} were used for training and testing.

	\begin{table*}[htp]
		% \vspace{-0.1cm}
		\setlength\tabcolsep{7.5pt}
		\centering
		% 		\small
		\caption{Comparison of pre-training methods fine-tuned on different downstream tasks with a frozen pre-trained model. Adapter$_{64}$ denotes the adapter-based pre-training with hidden size 64.}
		\begin{tabular}{l|c|c|cc|cc}
			\toprule
			\multirow{2}*{Method} & \multirow{2}*{\shortstack{Pre-trained params\\for SS and SE}} & \multicolumn{1}{c|}{SS} & \multicolumn{2}{|c|}{SE} & \multicolumn{2}{c}{ASR (WER)} \\
			\cline{3-7}
			&  & SI-SDR $\uparrow$ & PESQ $\uparrow$ & STOI $\uparrow$ & w/o $\downarrow$ & w/LM $\downarrow$  \\
			\midrule
			Fbank & -  & 9.15 & 1.84 & 87.25 & 26.36 & 17.37 \\
			Base HuBERT (Baseline)  & 94.70M$\times$0\% & \textbf{9.40} & \textbf{1.95} & \textbf{88.04} & \textbf{6.60} & \textbf{4.88}   \\
			\midrule
			\quad +Data Augmentation (960H$\times$4, D-A) & 94.70M$\times$100\% & 9.69 & 2.06 & 89.18 & 10.47 & 7.31 \\
			\quad \quad  +Primary-speaker denoising (PS-D)  & 94.70M$\times$100\% & 10.11  & 2.07 & 89.17 & \textbf{6.89} & \textbf{5.07}  \\
			\quad \quad  +Multi-label (ML)  & 95.02M$\times$100\% & \textbf{10.59} & \textbf{2.08} & \textbf{89.20} & 9.66 & 6.84  \\
			\midrule
			\qquad \quad  +Adapter$_{64}$ (A64) & 97.48M$\times$3.1\%  & 9.68 & 2.01 & 88.74 & 6.72 & 4.97  \\
			\qquad \quad +Adapter$_{128}$ (A128) & 99.84M$\times$5.5\%  & 9.77 & 2.02 & 88.85 & 6.64 & 4.94   \\
			\qquad \quad +Adapter$_{256}$ (A256) & 104.56M$\times$9.7\% & 10.17 & 2.04 & 89.04 & 6.80 & 5.02  \\
			\qquad \quad +Adapter$_{512}$ (A512) & 114.01M$\times$17.2\% & \textbf{10.19} & \textbf{2.05} & \textbf{89.10} & 6.87 & 5.06 \\
			\qquad \quad \ \ 	Inference w/o Adapter & - & - & - & - & \textbf{6.59} & \textbf{4.83}   \\
			\bottomrule
		\end{tabular}
		\vspace{-0.5cm}
		\label{table1}
	\end{table*}
	
	\subsection{Model and Training Setup}
	The HuBERT Base \cite{hubert} was selected as the baseline and the backbone. The data-augmentation pre-training (D-A) \cite{robustwav2vec} and primary-speaker denoising pre-training (PS-D) \cite{wavlm} were employed as the comparison methods. 
	
	\textbf{Backbone.} The convolutional module consists of seven 512-channel layers with strides $\left[5,2,2,2,2,2,2\right]$ and kernels $\left[10,3,3,3,3,2,2\right]$. The Transformer contains 12 layers with 768 dimensions, 3072 inner dimensions, and 12 attention heads. Moreover, the encoder-post module utilizes the FC and code embedding layer to predict masked speech codes. 
	
	The HuBERT of D-A, PS-D, and our methods were initialized with the official HuBERT. Then the D-A, PS-D, and multi-label HuBERT (ML) without adapters were pre-trained for 450K steps on 8 GPUs with a batch size of 150 seconds samples per GPU. The multi-label adapter-based HuBERT was pre-trained for 337.5K steps with a batch size of 200 seconds samples per GPU. Adam optimizer was used with the warm-up learning rate, which ramped up linearly from 0 to 3e-4 for the first 8\% steps and then decayed linearly to 0.
	
	For SS, a 3-layer 896-cell BiLSTM \cite{bilstm} with the 2-speaker ideal non-negative phase sensitive mask \cite{mask} was adopted and optimized by the permutation-invariant \cite{pit} based mean square error. For SE, a same BiLSTM as SS with the complex mask \cite{dccrn} was employed and optimized by the scale-invariant SNR \cite{deepcluster}. For ASR, a 2-layer BiLSTM with 1024 cells was used and optimized by the character-unit CTC loss. The SUPERB-SG-style fine-tuning \cite{superbsg} was firstly executed for three tasks. And then, we fine-tuned the pre-trained and downstream models together for SS and SE, respectively.
	%  (86 epochs for SS and SE, 112 epochs for ASR, lr 1e-4)
	% (86 epochs, lr 1e-4)

	\subsection{SUPERB-SG-style Fine-tuning}
	Following the SUPERB-SG, the fine-tuning with the frozen pre-trained model was executed to evaluate the model's universality on SS, SE, and ASR. The outputs of all Transformer layers were weighted with learnable weights and summed as the input to the downstream model. The evaluation metrics for SS and SE are the scale-invariant signal-to-distortion ratio (SI-SDR) \cite{deepcluster}, perceptual evaluation of speech quality (PESQ) \cite{pesq}, and short-time objective intelligibility (STOI) \cite{stoi}. The word error rate (WER) is counted for ASR.
	
	As shown in Table~\ref{table1}, compared to the Fbank, significant improvements in ASR but limited ones on SS and SE are brought by the original HuBERT. Based on HuBERT, the target domain data augmentation and PS-D improve the performance on SS and SE. But the absence of multi-speaker information limits their improvements on SS. Unsurprisingly, our proposed multi-label pre-training achieves the best results on SS and SE (SI-SDR:10.59/PESQ:2.08/STOI:89.20), which can retain multiple-speaker information while extracting targets from a mixture. However, compared to the original HuBERT, the WER on ASR of multi-label pre-training relatively increases by 40\%. We speculate that multiple-information retaining is adverse to single-speaker ASR.
	
	The degradation of the original tasks is undesired. As can be seen from the results, based on multi-label pre-training, the adapter-based methods with a few pre-trained parameters achieve comparable performance to the full pre-training on SS and SE, and the degradation on ASR is substantially suppressed. We speculate that training only a few additional parameters for a new task can reduce the bias introduced by the new task. In this way, the model can leverage its prior knowledge to learn a new task. As expected, inferencing without adapters performs the same as the original HuBERT on ASR. 
	
	\subsection{Fine-tuning for SS or SE}
	The SSL models were frozen and operated in a constrained condition in previous experiments. Fine-tuning experiments with the unfrozen SSL models for SS and SE were conducted to evaluate our proposed method under a more realistic condition. As shown in Table~\ref{allfinetune}, the overall results are consistent with Table~\ref{table1}, which shows that multi-label pre-training performs excellently on SS and SE. As the adapter size increases, the performance of adapter-based multi-label pre-training gradually approaches the performance of full pre-training.
	
	\begin{table}[htp]
		\vspace{-0.2cm}
		\setlength\tabcolsep{2.7pt}
		\centering
		\small
		\caption{Performance of different methods fine-tuned with unfrozen SSL model on SS or SE.}
		\begin{tabular}{lcccccccc}
			\toprule
			Task  & Baseline & D-A & PS-D & ML & A64 & A128 & A256 & A512\\
			\midrule
			SE$_\text{PESQ}$ & 2.11 & 2.21 & \textbf{2.26} & 2.25 & 2.16 & 2.18 & 2.21 & 2.22 \\
			SS$_\text{SI\-SDR}$ & 10.71 & 10.79 & 11.00 & \textbf{11.26} & 10.77 & 10.79 & 10.99 & 11.06 \\
			\bottomrule
		\end{tabular}
		\vspace{-0.3cm}
		\label{allfinetune}
	\end{table}
	
	\subsection{Consumed Time for Pre-training}
	The consumed time per epoch of different methods was counted, as shown in Table~\ref{time}. The adapter-based pre-training is 19\% relative faster than the full pre-training. We believe that the adapter-based pre-training may be a practical choice to evaluate a pre-training method, which converges faster with less GPU memory (less trainable parameters) and performs consistently with full pre-training. 
	
	\begin{table}[htp]
		\centering
		\vspace{-0.4cm}
		\small
		\caption{Consumed time per epoch of full pre-training and adapter-based pre-training (on the augmented dataset).}
		\begin{tabular}{cccccc}
			\toprule
			& ML & A64 & A128 & A256 & A512\\
			\midrule
			Time (min)& 162 & 131 & 131 & 132 & 133\\
			\bottomrule
		\end{tabular}
		\vspace{-0.6cm}
		\label{time}
	\end{table}
	
	\section{conclution}
	
	We believe that the data domain mismatch and task mismatch are the culprits for the mediocre performance of SSL features on SS and SE. In this study, we propose a multi-label pre-training scheme to introduce separation and denoising into the masked prediction loss of HuBERT, and its significant improvements on SS and SE are experimentally verified. However, the differences in task requirements degrade its performance on other tasks, such as ASR. Therefore, the lightweight adapters are further inserted into HuBERT's Transformer to be pre-trained for SS and SE. In this way, comparable improvements on SS and SE can be achieved while retaining outstanding performance on the original tasks.
	
	\ninept
	\bibliographystyle{IEEEbib}
	\bibliography{strings,refs}
	
\end{document}